\documentclass[11pt,a4paper]{article}
\usepackage{setspace}
\usepackage{graphicx}
\begin{document}
%\begin{center}
\title{\bf\large The $\phi(1020)\to\pi^0\pi^0\gamma$ decay\thanks{
The work is partially supported by RFBR 
(Grants No 99-02-16813, 99-02-16815, 00-02-17478, 00-02-17481) and
``Russian Universities'' (Grant No 3H-339-00).}  }
\author{
M.N.~Achasov, 
S.E.~Baru, 
K.I.~Beloborodov, 
A.V.~Berdyugin, 
A.V.~Bozhenok,\\ 
A.D.~Bukin, 
D.A.~Bukin, 
S.V.~Burdin, 
T.V.~Dimova,
A.A.~Drozdetsky, \\  
V.P.~Druzhinin, 
M.S.~Dubrovin,
I.A.~Gaponenko,
V.B.~Golubev, 
\underline{V.N.~Ivanchenko}\thanks{email:V.N.Ivanchenko@inp.nsk.su},\\
  A.A.~Korol, 
I.A.~Koop, 
S.V.~Koshuba, 
G.A.~Kukartsev, 
A.A.~Mamutkin,\\
E.V.~Pakhtusova, 
A.A.~Salnikov, 
S.I.~Serednyakov, 
V.V.~Shary, 
Yu.M.~Shatunov,\\ 
V.A.~Sidorov, 
Z.K.~Silagadze, 
A.N.~Skrinsky,
A.A.~Valishev,
A.V.~Vasiljev \\  \\
Budker Institute of Nuclear Physics, Novosibirsk, 630090, Russia\\
 }
\date{}
\maketitle

%\end{center}

\begin{abstract}
In the SND experiment at VEPP-2M  $e^+e^-$ collider the
$\phi(1020)\to\pi^0\pi^0\gamma$ decay was studied
and its branching ratio was measured:
$B(\phi\to\pi^0\pi^0\gamma)=(1.221\pm 0.098\pm0.061)\cdot10^{-4}$.
It was shown, that $f_0(980)\gamma$ intermediate state dominates
in this decay and
the $f_0(980)$-meson parameters were obtained. 
\end{abstract}

{\it PACS:}{12.39.Mk, 13.40.Hq, 14.40.Cs}

{\it Keywords:} scalar, vector, decay, four-quark, gamma, radiative

%\twocolumn

%\nopagebreak

\section{Introduction}

The first work to search for the decay $\phi\to\pi^0\pi^0\gamma$
was performed in 1987 \cite{IVN1,REP}.
The observation of this decay based on
analysis of the data of  the PHI96 experiment  \cite{PHI96}
was reported in 1997 \cite{PPG1,PPG2} by the SND detector \cite{SND}.
Observed decay rate of about $10^{-4}$ was unexpectedly high and
it was also shown that $f_0(980)\gamma$ intermediate state
dominates in this decay.
Later these results were confirmed
by the CMD-2 experiment \cite{PPGCMD2}. 
 
In this work results  are presented of a new
experimental study of the process 
\begin{equation}
e^+e^-\to\phi\to\pi^0\pi^0\gamma
\label{EQ0}
\end{equation}
based on complete SND 
statistics of about $2\cdot 10^7$ 
produced $\phi$ mesons.
The experiments were carried out at 
VEPP-2M $e^+e^-$ collider in Novosibirsk.
 The total integrated luminosity collected in PHI96 \cite{PHI96}
and PHI98 \cite{PHI98} experiments was  
$4~pb^{-1}$ in 20 energy points
and $8~pb^{-1}$ in the 16 energy points respectively.

\section{Data analysis}

Main resonant background to the reaction  (\ref{EQ0})
comes from the process
\begin{equation}
{e^+e^-\to\phi\to\eta\gamma\to3\pi^0\gamma}
\label{EQ1}
\end{equation}
due to the merging of
photons and/or loss of photons through the openings in the
calorimeter. 
A smaller background comes from the decay mode \cite{A0G}
\begin{equation}
{e^+e^-\to\phi\to\eta\pi^0\gamma}
\label{EQ1A}
\end{equation}
Copiously produced $\phi\to K_S K_L \to\pi^0\pi^0K_L$ events
may contribute to the background but 
this decay is almost completely rejected by our cuts.
The main source of non-resonant background is the process
\begin{equation}
e^+e^-\to\omega\pi^0\to\pi^0\pi^0\gamma.
\label{EQ3}
\end{equation}
The QED background from $e^+e^-$ annihilation into $4,5\gamma$ was
found  negligible.

The data analysis was described in detail in our previous
publications \cite{PPG1,PPG2}. In this work basically
the same selection
cuts were used requiring five isolated photons in the calorimeter
with good shower profiles and  event kinematics
consistent with the $\pi^0\pi^0\gamma$ hypothesis (93~\% c.l.).
To suppress spurious photons from beams background
all photons with
energies less than $60~\mbox{MeV}$ at angles with respect to the beams
less than 36 degrees were excluded from further analysis.
The rate of events with spurious photons
was about 4~\% in PHI96 experiment and  8~\% in PHI98. 
To account for this effect during processing of simulated events
spurious hits, taken from the experiment, were admixed to Monte Carlo
ones.

\begin{figure} % fig 6
\centerline{\includegraphics[width=0.7\textwidth]{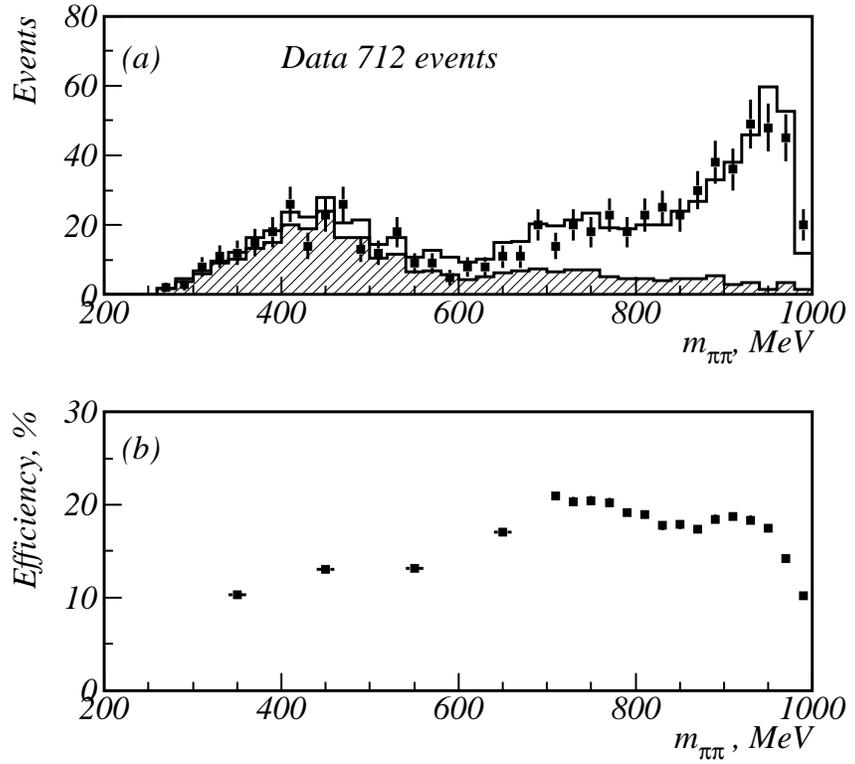}}
\caption{The mass spectra for $\pi^0\pi^0\gamma$ events:
a --- the invariant mass distribution
of $\pi^0\pi^0$ pairs 
without acceptance corrections,
beam energy $1015~\mbox{MeV} < 2E < 1025~\mbox{MeV}$,
histogram --- data, 
shaded histogram --- estimated background;
b --- the detection efficiency for  the process
$e^+e^-\to S\gamma\to\pi^0\pi^0\gamma$, S is a scalar state.
}
\label{fig6}
\end{figure}

After applying cuts the sample of the $\pi^0\pi^0\gamma$ candidates  was 
selected  (Fig.\ref{fig6}a).
The main background is due to the process (\ref{EQ3}) 
but processes (\ref{EQ1}) and (\ref{EQ1A}) contribute to this 
event sample also. The estimated number of events
of the reaction (\ref{EQ0}) in Fig.\ref{fig6}a is 
 $419\pm 31$.
The detection efficiency (Fig.\ref{fig6}b)
for the process (\ref{EQ0}) as a function of $\pi^0\pi^0$ invariant mass
was determined using simulation 
of the process   $\phi\to S \gamma \to \pi^0 \pi^0\gamma$, where $S$
is a scalar state with a mass ranging from 300 to $1000~\mbox{MeV}$.
The whole mass range was divided into 19 bins and within each bin
simulation with uniform  $m_{\pi\pi}$ distribution was performed.
This simulation was used for determination of the $\pi^0\pi^0$
invariant mass resolution and event misidentification probabilities
as well.

\begin{figure} % fig 4
\centerline{\includegraphics[width=0.9\textwidth]{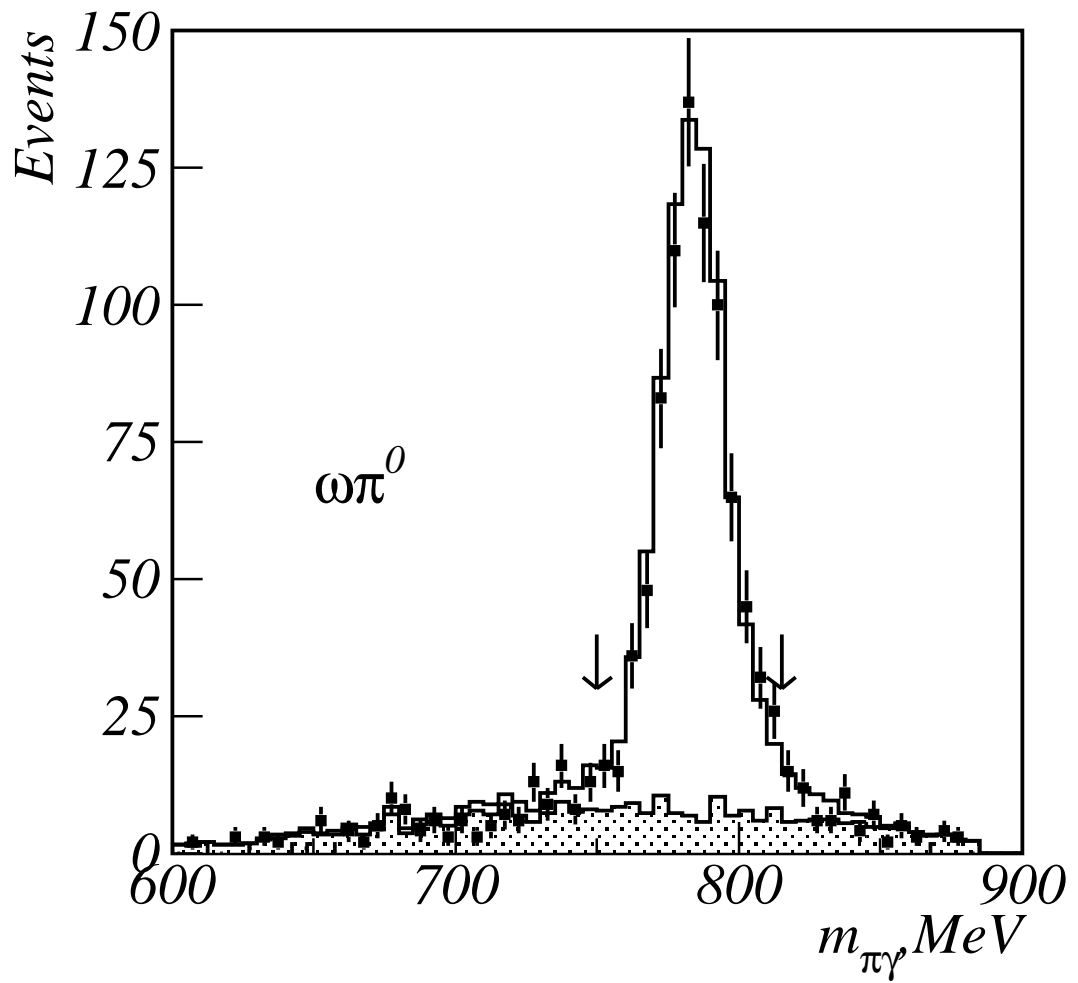}}
\caption{Distribution of $\pi^0\gamma$ invariant mass for
$m_{\pi\pi}<700~\mbox{MeV}$. 
Points --- data, histogram --- simulation, 
shaded histogram ---
sum of contributions from 
the $\phi$ decays,
arrows --- selection of $\omega\pi^0$.}
\label{fig4}
\end{figure}

For the background process
(\ref{EQ3})
 the $\pi^0\gamma$ invariant mass distribution is peaked at
$\omega$-meson mass.
The $m_{\pi\gamma}$ parameter is an invariant mass of the
recoil photon and one of $\pi^0$ mesons, closest 
to the $\omega$-meson mass (Fig.\ref{fig4}).
Events with
\begin{equation}
750~\mbox{MeV} < m_{\pi\gamma} < 815~\mbox{MeV}
\label{OP}
\end{equation}
are mainly ones of the process (\ref{EQ3}).
Moreover in the energy range of the experiment
$\pi^0\pi^0$ invariant mass in the process
(\ref{EQ3}) must be lower than
$700~\mbox{MeV}$. For further analysis 
the selected $\pi^0\pi^0\gamma$ events
were divided into three classes: 
\begin{enumerate}
\item
$m_{\pi\pi} < 700~\mbox{MeV}$ and not (\ref{OP});
\item
$m_{\pi\pi} > 700~\mbox{MeV}$ and not (\ref{OP});
\item
$m_{\pi\pi} < 700~\mbox{MeV}$ and  (\ref{OP}).
\end{enumerate}
In the second class the $f_0\gamma$ mechanism 
dominates, in the third --- the process (\ref{EQ3}). 

\begin{figure} % fig 5
\centerline{\includegraphics[width=0.9\textwidth]{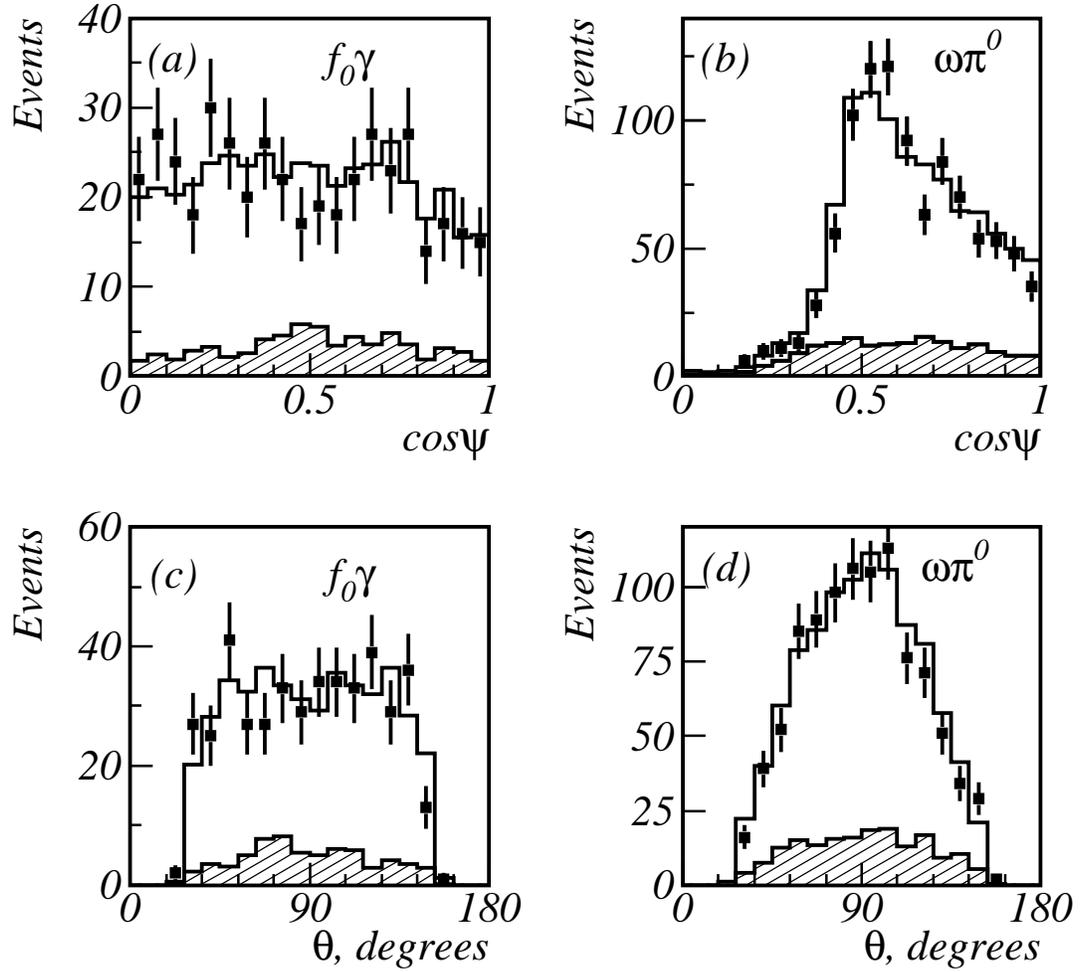}}
\caption{The angular spectra for $\pi^0\pi^0\gamma$ events:
a, b --- distribution of cosine of $\psi$,
the angle between directions of $\pi^0$ and recoil $\gamma$
in the rest frame of $\pi^0\pi^0$ system;
c, d ---  distributions of $\theta$, angle of the
recoil $\gamma$ with respect to the beam.
Points --- data, histogram --- simulation,
shaded histogram ---
sum of contributions from 
background processes for the reaction (\ref{EQ0}) (a,c)
and for (\ref{EQ3}) (b,d).
Acceptance corrections are not applied.
Beam energy is $1015~\mbox{MeV} < 2E < 1025~\mbox{MeV}$.}
\label{fig5}
\end{figure}

For  events of the second and third classes
the experimental and simulated
distributions in  $\psi$ and $\theta$ angles were compared.
The $\psi$ is an angle of the recoil photon with respect to pion
direction in the $\pi^0\pi^0$ center of mass reference frame.
The $\theta$ is an angle between recoil photon and the beam.
It can be seen in Fig.\ref{fig5}a,c 
that in the second class the
distribution well matches one expected for
the intermediate scalar state.
Corresponding distributions in the third class
agree with ones expected for the intermediate 
$\omega\pi^0$ state (Fig.\ref{fig5}b,d).

\section{Fitting of cross sections}

The visible cross sections $\sigma_i(s)$ 
were measured for each energy point in both PHI96 and PHI98 experiments,
here $s = 4E^2$ and  $i$ is a class number ($i$=1,2,3) defined above.
These cross sections 
were fitted according to the following formulae:
\begin{equation}
\sigma_i(s)  =  \sum_{j=1,2}{
\alpha^j_{\pi\pi\gamma}(s)\epsilon^{ij}_{\pi\pi\gamma}\sigma^j_{\pi\pi\gamma}(s)}
 +  \alpha_{\eta\gamma}(s)\epsilon^i_{\eta\gamma}\sigma_{\eta\gamma}(s)
 +  
\alpha_{\eta\pi\gamma}(s)\epsilon^i_{\eta\pi\gamma}\sigma_{\eta\pi\gamma}(s)
 + 
\alpha_{\omega\pi}(s)\epsilon^i_{\omega\pi}\sigma_{\omega\pi}(s),
\label{EQSIG}
\end{equation}
where  $\sigma^j_{\pi\pi\gamma}(s)$ is a cross section
for the $j$-th $m_{\pi\pi}$ interval ($j=1$ for $m_{\pi\pi}<700~\mbox{MeV}$,
$j=2$ for $m_{\pi\pi}>700~\mbox{MeV}$),
$\epsilon^{ij}_{\pi\pi\gamma}$ are probabilities 
for the $\pi^0\pi^0\gamma$ 
events belonging to $j$-th $m_{\pi\pi}$ interval to be detected
as the class $i$ events
(Tabl.\ref{T96},\ref{T98}),
 $\sigma_f(s)$  are production cross sections
of the final state $f$
($f =  \eta\gamma$, $\eta\pi^0\gamma$,
$\omega\pi^0$),
$\epsilon^i_f$
are detection efficiencies of the state $f$ in the class $i$,
$\alpha_{f}(s)$ --- radiative corrections
and corrections to the energy spread in the beam \cite{DB}.
The expression for the $\sigma_{\eta\gamma}(s)$ was taken from  
Ref.\cite{ETGPG},
the values of the $\sigma_{\omega\pi}(s)$ as well as 
$\alpha_{\omega\pi}(s)$
were taken from 
our  previous work \cite{OMPI}.
The efficiencies $\epsilon^{ij}_{\pi\pi\gamma}$ were
estimated using the experimental 
mass spectrum and  the simulation of $S\gamma$ events
described in the previous section.
For the cross sections
$\sigma_{f}(s)$ (f is $\pi^0\pi^0\gamma$ or $\pi^0\eta\gamma$)
the following expression was used:
\begin{eqnarray}
\sigma_{f}(s) & = & \frac{12\pi \Gamma^2_{\phi}B(\phi\to ee)
B(\phi \to R\gamma) m^3_{\phi}F(s)}
{s^{3/2} \left | D_{\phi}(s) \right |^2},  \nonumber \\
F(s) & = & \frac{\Gamma_{\phi R\gamma}(s)}
{\Gamma_{\phi R\gamma}(m^2_{\phi})}
,  \nonumber \: \:
R = f_0, a_0. 
\label{EQPPG}
\end{eqnarray}
Here the inverse propagator of $\phi$ was written according to
Ref.\cite{FLOOK}:
$D_{\phi}(s) = m^2_{\phi} - s -i \cdot \sqrt{s}\Gamma_{\phi}(s)$ and the
energy dependence of partial widths $\Gamma_{\phi R\gamma}(s)$ was taken from 
Ref.\cite{IVN2}. The branching ratio of the decay
(\ref{EQ1A}) was obtained in SND work \cite{A0G}.

Note, that 
there is a dip
in the cross section (\ref{EQ3}) at the $\phi$-meson mass,
due to  destructive
interference between the main non-resonance process, the  
$\phi\to\omega\pi^0$ decay, and the process 
\begin{equation}
e^+e^-\to\phi\to\rho\pi^0\to\pi^0\pi^0\gamma .
\label{EQ2}
\end{equation}

To check the accuracy of the detection efficiency, 
obtained by simulation, fully reconstructed
events of the process (\ref{EQ1}) with 7 photons
in the final state were analyzed.
Since this process is similar to (\ref{EQ0}), the energy,
invariant mass, and angular cuts were the same as in 
$\pi^0\pi^0\gamma$ analysis, while threshold values for
kinematic fitting parameters were scaled to provide
same confidence level.
Obtained branching ratio of the decay (\ref{EQ1})
was smaller than that from the SND experiment \cite{ETGPG}
$B(\phi\to\eta\gamma)=(1.338\pm0.053)\%$
by a factor of $\xi=0.89\pm 0.04$. This value  was
taken as a correction factor for both effect 
and background.
This correction accounts for imprecise simulation
of tails of distributions of parameters used in 
the cuts.

\begin{figure} % fig 7
\centerline{\includegraphics[width=0.9\textwidth]{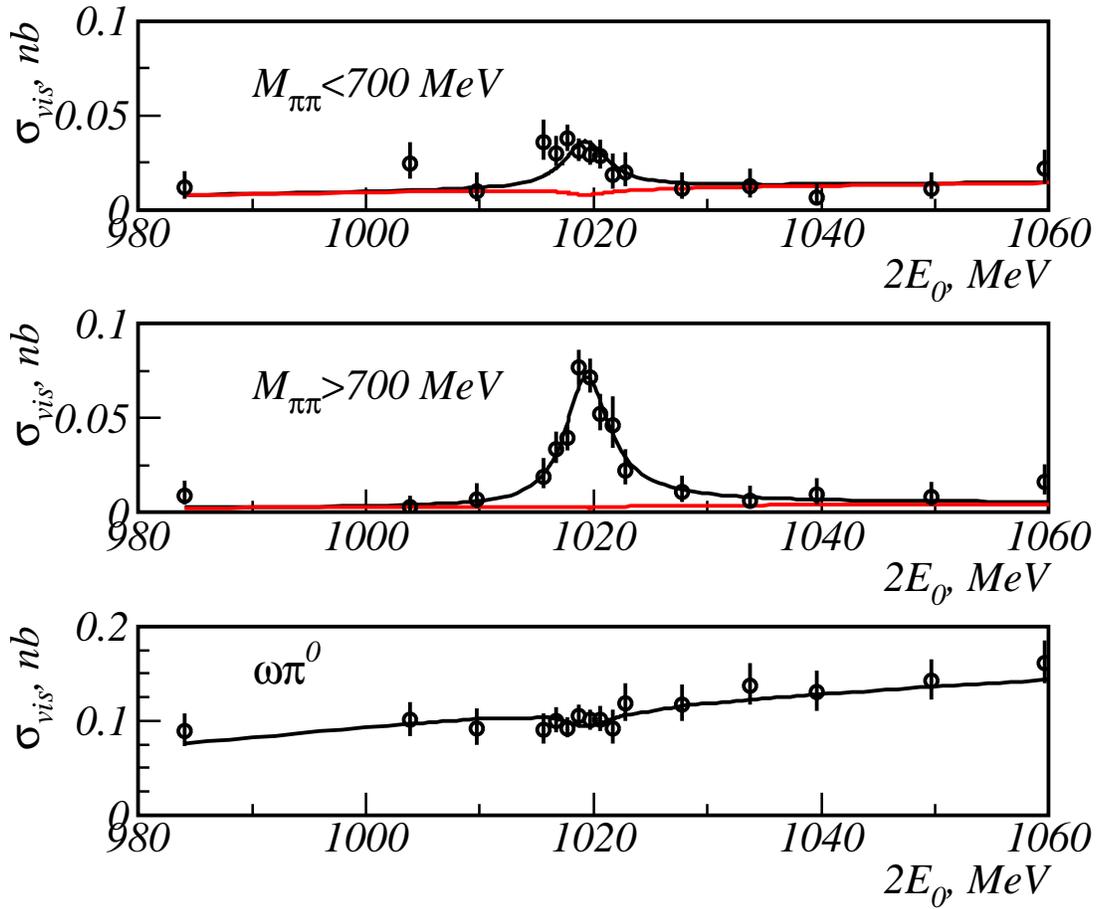}}
\caption{ Energy dependence of the visible
$e^+e^-\to\pi^0\pi^0\gamma$ cross section for the three classes
of events described in the text for the PHI98 experiment.
Points --- data, solid line --- fit, 
dotted line --- sum of background contributions to the decay
$\phi\to\pi^0\pi^0\gamma$.}
\label{figsigma}
\end{figure}

Free parameters of the fit were the branching ratios in the 
$\phi$-meson peak $B(\phi\to \pi^0\pi^0\gamma)$ in two intervals
of $m_{\pi\pi}$ (Tabl.\ref{T96},\ref{T98}).
Uncertainties in the values of $\Gamma_{\phi}$, detection
efficiencies, integrated luminosity, and energy spread of the beam
were taken into account. The value of $B(\phi\to\eta\gamma)$
was fixed to its SND value, so
our results are normalized to the branching
ratio of the decay $\phi\to\eta\gamma$. 
The fitting results  for both 
experiments are shown in the Table~\ref{TF},
corresponding curves for PHI98
are shown in Fig.\ref{figsigma}.
Because the most of  
$\pi^0\pi^0\gamma$ events concentrate 
at high $m_{\pi\pi}$ similar analysis 
was performed 
for the mass range $m_{\pi\pi} > 900~\mbox{MeV}$.

\section{Results}

Because the fitting results for the
PHI96 and PHI98 experiments are in a good agreement
with each other (Tabl.~\ref{TF}) they was
 combined:
\begin{eqnarray}
B(\phi\to\pi^0\pi^0\gamma)  =  (1.034\pm 0.066\pm0.046)\cdot 10^{-4}, 
\;  
m_{\pi\pi}  >  700~\mbox{MeV}, 
\label{EQ60} \\
B(\phi\to\pi^0\pi^0\gamma)  =  (0.559\pm 0.053\pm0.025)\cdot 10^{-4},
\;
 m_{\pi\pi}  >  900~\mbox{MeV},
\label{EQ6} \\
B(\phi\to\pi^0\pi^0\gamma)  =  (0.124\pm 0.065\pm 0.006)\cdot 10^{-4}, 
\;
 m_{\pi\pi}  <  700~\mbox{MeV}.
\label{EQ61}
\end{eqnarray}
In (\ref{EQ60}--\ref{EQ61}) the
first error is statistical and the second one is systematic,
which is  determined mainly by current accuracy of
$B(\phi\to\eta\gamma)$ (4~\%) and model uncertainty  (2~\%) of
$F(s)$ in the formula (\ref{EQPPG}). 
From the data (\ref{EQ60}) and (\ref{EQ61}) 
one can evaluate the total value
\begin{equation}
B(\phi\to\pi^0\pi^0\gamma)  =  (1.158\pm 0.093\pm0.052)\cdot 10^{-4},
\;
 \phi  \to  \omega\pi^0~excluded,
\label{EQ62}
\end{equation}
which is 
the total branching ratio excluding
$\phi\to\omega\pi^0$ contribution,
because the
kinematic region of this decay
is excluded by our cuts. It was studied in the SND
work \cite{OMPI0}, from which one can obtain 
the branching ratio to the $\omega\pi^0$ region
$B(\phi\to \pi^0\pi^0\gamma) = (0.063\pm 0.031)\cdot 10^{-4}$
 and add it to (\ref{EQ62}) to find
the total branching ratio
\begin{equation}
B(\phi\to\pi^0\pi^0\gamma) = (1.221\pm 0.098\pm0.061)\cdot 10^{-4}.
\label{EQ64}
\end{equation}
In this work
the branching ratios were measured relatively to 
$B(\phi\to\eta\gamma)$. In such approach some systematic errors
cancel allowing smaller total systematic uncertainties:
\begin{eqnarray}
\frac{B(\phi\to\pi^0\pi^0\gamma)}{B(\phi\to\eta\gamma)}
& = & (0.913\pm 0.073\pm0.029)\cdot 10^{-2}, 
\;
 total; 
\nonumber \\
\frac{B(\phi\to\pi^0\pi^0\gamma)}{B(\phi\to\eta\gamma)}
& = & (0.865\pm 0.070\pm0.017)\cdot 10^{-2},
\;
  \omega\pi^0~excluded;
\nonumber \\
\frac{B(\phi\to\pi^0\pi^0\gamma)}{B(\phi\to\eta\gamma)}
&=&(0.773\pm 0.049\pm0.016)\cdot 10^{-2},
\;
   m_{\pi\pi}  >  700~\mbox{MeV}; 
\nonumber \\
\frac{B(\phi\to\pi^0\pi^0\gamma)}{B(\phi\to\eta\gamma)}
&=&(0.418\pm 0.040\pm0.009)\cdot 10^{-2},
\;
 m_{\pi\pi}  >  900~\mbox{MeV}.
\label{EQ84}
\end{eqnarray}

To investigate the dynamics of the reaction (\ref{EQ0})
the  $\pi^0\pi^0$ invariant mass spectrum (Fig.\ref{fig6}a)
was studied in a narrower energy interval 
$2E=(1015-1025)~\mbox{MeV}$ containing almost all $\phi$-meson 
events and smaller $\omega\pi^0$ background. 
The estimated background from the processes
(\ref{EQ1}--\ref{EQ3}) was subtracted using SND
data \cite{A0G,OMPI,ETGPG}.
Resulting raw mass spectrum is distorted due to finite 
detector resolution and misidentification of recoil photon 
in some $\pi^0\pi^0\gamma$ events.
In order to obtain corrected  spectrum
the matrix   $\epsilon_{i,j}$ was constructed, where each $\epsilon_{i,j}$
is a  probability to get a
reconstructed $m_{\pi\pi}$ value within the invariant mass
bin $i$ for an  simulated event with actual $m_{\pi\pi}$
 uniformly distributed in the bin $j$.
Note, that
in the two highest  bins  the simulated mass distribution 
were subdivided 
into smaller bins to take into account rapid decrease
of the mass spectrum.
The diagonal elements of the matrix
and their immediate neighbors describe detector resolution.
They
turned out to be
 much larger than others which are due to misidentification
of the recoil photon.
Using this matrix and the experimental mass spectrum
the  number of events with  misidentification were 
estimated and subtracted from each bin. 
This procedure does not affect the diagonal elements 
of the matrix and their immediate neighbors but zeroes 
all others.

After normalization to the total integrated 
luminosity
and detection efficiency the differential branching
ratios $S_{exp}(m_i)=dB(\phi\to \pi^0\pi^0\gamma) / dm_{\pi\pi}$ for both
experiments were
obtained. They are in a good agreement and thus can be combined
together (Tabl.\ref{T0}).
Using the matrix of efficiencies it is possible to restore
the physical mass spectrum but such a procedure
makes the values of corrected spectrum
strongly correlated. To compare these results 
with model predictions one has to deal with
a cumbersome covariance matrix.
To simplify the analysis we instead transformed bare
theory predictions $S_{th}(m)$ 
using the values of efficiencies from Table~\ref{T0}:
\begin{equation}
S_{cor}(m_i)  =  \epsilon_{i,i-1}S_{th}(m_{i-1}) 
  +  \epsilon_{i,i}S_{th}(m_{i}) 
  +  \epsilon_{i,i+1}S_{th}(m_{i+1}),
\label{ENCAP}
\end{equation}
here $S_{cor}(m_i)$ is a corrected for the detector response
theoretical mass spectrum, 
which should be used for approximation of the
experimental spectrum $S_{exp}(m_i)$   by 
the least squares method. 
This approach is used in our fits, which are 
discussed in the next section.

\section{Discussion}

\begin{figure} % fig 8
\centerline{\includegraphics[width=0.7\textwidth]{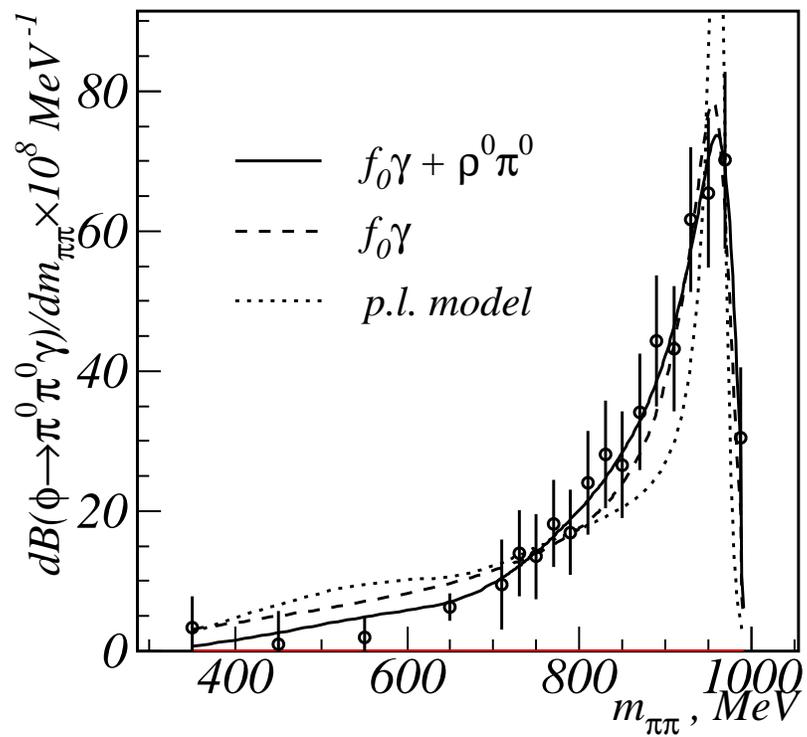}}
\caption{The  $\pi^0\pi^0$ invariant mass spectrum after background
subtraction and acceptance corrections:
points --- data, solid line -- the result of the  fit,
dashed line --- the result of the fit without $\rho^0\pi^0$
contribution;
dotted line --- the result of the fit in the point-like 
transition model.}
\label{fig8}
\end{figure}

The $\pi^0\pi^0$
invariant mass spectrum (Fig.\ref{fig8})
was approximated by 
the following function: 
\begin{equation}
S_{th}(m)  =  S_{S\gamma}(m) +   
S_{\rho\pi}(m)  +  2\sqrt{S_{S\gamma}(m)S_{\rho\pi}(m)}\cos\theta, 
\label{INT}
\end{equation}
where $S_{S\gamma}(m)$ is the contribution of the $f_0\gamma$
and $\sigma\gamma$ 
decays,
$S_{\rho\pi}(m)$ is the contribution of the process (\ref{EQ2}), 
and $\theta$ is an interference phase. This formula is exact only in case
of identical angular distributions in $S\gamma$ and $\rho^0\pi^0$
intermediate states.
Provided small $\rho^0\pi^0$ contribution, 
the difference in the angular
distributions for the $\pi^0\pi^0\gamma$ final state
can be approximately accounted for by fitting of $\theta$.
The $S_{\rho\pi}(m)$ dependence was taken from
\cite{IVN2,IVN5}. Corresponding  branching ratio 
is $B(\phi\to\rho\pi\to\pi^0\pi^0\gamma)=(1.2\pm 0.2)\cdot 10^{-5}$
\cite{IVN5}.
For $S_{S\gamma}(m)$ the expression
from Ref.\cite{IVN2} was used
\begin{eqnarray}
S_{S\gamma}(m) & = &
\frac {2m^{2} \Gamma_{\phi f\gamma}
\Gamma_{f\pi^0\pi^0}}{\pi\Gamma_{\phi}}  
    \left |  \frac {1}{D_f(m)}  + 
\frac{A} {D_{\sigma}(m)}
\right |^2,                                   \nonumber \\
\Gamma_{f\pi^0\pi^0} & = & 
\frac{g^2_{f\pi^+\pi^-}}{32\pi m} \sqrt{1-\frac{4m^2_{\pi}}{m^2}},
\label{SPEC}
\end{eqnarray}
where 
$D_S(m) = m^2_S - m^2 + \Pi_S(m) -i \cdot m\Gamma_S(m)$ 
is  $f_0$ or $\sigma$  inverse propagator, $\Pi_S(m)$ 
takes into
account finite width corrections \cite{IVN2},
$A$ is a complex parameter taking into account difference in coupling
constants of $f_0$ and $\sigma$ mesons.

In the kaon loop model of the $\phi\to f_0 \gamma$ transition \cite{IVN2} 
$\Gamma_{\phi f\gamma}$ is proportional
to the product  $g^2_{fK^+K^-}g^2_{f\pi^+\pi^-}$. 
The dependence of $\Gamma_{\phi f\gamma}$ on
$m$ in this model is different from the model with the point-like 
$\phi\to f_0\gamma$ transition and can be interpreted
as a form factor, which stops the
$\Gamma_{\phi f\gamma}$ growth at higher photon energies.

The free parameters of the fit were $\theta$, $A$, $m_f$,
and coupling constants $g_{fK^+K^-}$ and $g_{f\pi^+\pi^-}$ while
the width and mass of the $\sigma$ meson were fixed at 
$m_{\sigma}=600~\mbox{MeV}$ and
$\Gamma_{\sigma}=400~\mbox{MeV}$ \cite{ISHIDA}.
For $\sigma$ meson the Breit-Wigner
propagator with $\Pi_{\sigma}(m)=0$ and $\Gamma_{\sigma}(m)=const$ was used.
The fitting result is  
$A=-0.065^{+0.083}_{-0.167}$,
indicating  that
$\sigma\gamma$ contribution is compatible with zero.
This allows to reduce the number of free fitting parameters by
fixing  
$A=0$. For the remaining free parameters
the following values were obtained:
\begin{eqnarray}
m_f & = & (969.8\pm4.5)~\mbox{MeV}, \nonumber \\
g^2_{fK^+K^-}/4\pi & = & 2.47^{+0.73}_{-0.51}~\mbox{GeV}^2, \nonumber \\
g^2_{f\pi^+\pi^-}/4\pi & = & 0.54^{+0.09}_{-0.08}~\mbox{GeV}^2,\nonumber \\
g^2_{fK^+K^-}/g^2_{f\pi^+\pi^-} & = & 4.6\pm0.8, \nonumber \\
\theta & = & (180\pm 36)~degrees, \nonumber \\
\chi^2/n_D & = & 3/14.
\label{FITS}
\end{eqnarray}
Note, that although  contribution from the process (\ref{EQ2})
is small, and fitting with this contribution set to zero still gives
high  confidence level ($\chi^2/n_D=10/14$), but 
corresponding fitting curve (fig.\ref{fig8}) shows visible 
systematic deviations from the observed mass spectrum.

The fitting results (\ref{FITS}) demonstrate that the $f_0\gamma$
mechanism dominates in the decay (\ref{EQ0}),
and contributions from $\sigma\gamma$ and
$\rho^0\pi^0$ are small. Neglecting these contributions and
 assuming natural isotopic ratio
$B(f_0\to\pi^+\pi^-)=2B(f_0\to\pi^0\pi^0)$
one can obtain from (\ref{EQ62}) 
\begin{eqnarray}
B(\phi\to f_0\gamma) & = & (3.5\pm 0.3_{-0.5}^{+1.3})\cdot 10^{-4},
\nonumber \\
\frac{B(\phi\to f_0\gamma)}{B(\phi\to\eta\gamma)}
 & = & (2.6\pm 0.2_{-0.3}^{+0.8})\cdot 10^{-2}.
\label{EQ86}
\end{eqnarray}
where the second errors are systematic which includes uncertainty in
the interference terms. 
These errors were estimated by 
summation of systematic uncertainty from (\ref{EQ62})
and the difference between (\ref{EQ86})  and the value 
$B(\phi\to f_0\gamma)=4.6\cdot 10^{-4}$
calculated from  (\ref{FITS}) with $\sigma\gamma$ and
$\rho^0\pi^0$ contributions taken into account.

The $f_0$-meson width in the model \cite{IVN2} is a rapid function of $m$.
At the resonance pole it is equal to  $\Gamma_f(m_f) =
-Im(D_f(m_f))/m_f= (201\pm 28)~\mbox{MeV}$.
Due to
its close proximity to $K\overline{K}$ threshold this value
is significantly larger than
visible in Fig.\ref{fig8} $f_0$ width $\sim 100~\mbox{MeV}$.
Note, that
if the width corrections to the real part of $f_0$ propagator
are set to zero the values of fitting parameters
including $f_0$ mass and width change significantly:
$m_f  =  985^{+16}_{-12}~\mbox{MeV}$, 
$\Gamma_f  =  122\pm 13~\mbox{MeV}$,
but such assumption looks much less attractive from
the theoretical point of view. The model dependence 
of the $f_0$ mass and width  explains disagreement
between (\ref{FITS}) and some data from PDG
\cite{PDG98}.

The model of Ref.\cite{IVN2} which takes into
account the kaon loop contribution
in the decay under study fits well the experimental spectrum. 
On the contrary, it is
not possible to describe the data 
using the model with point-like transition mechanism,
in which the transition form factor is equal to unity and
$\Gamma_{\phi f_0\gamma}$ is proportional to $\omega^3$,
where $\omega$ is the photon energy. In this case
the fitting curve does not 
follow the experimental spectrum
and the $\chi^2/n_D = 28/14$ (fig.\ref{fig8}).
So, 
kaon loop contribution, which provides non point-like 
transition $\phi\to f_0\gamma$ is confirmed by the our data.

The values of coupling constants (\ref{FITS})
are in a good agreement with the 4-quark MIT bag
model predictions \cite{IVN2,IVN13,ACHAS3}. 
New theoretical studies \cite{NEW1,NEW2,NEW3,NEW4,NEW5} 
were performed after first  observation
in order to explain the data. It was shown that
it is possible to describe  experimental results 
using different theoretical approaches. 
In these models the 4-quark component exists directly or 
``de facto'' as a result of the kaon loops contribution. 

\section{Conclusion}

Results of this work
confirming our first publications \cite{PPG1,PPG2}
and CMD-2 measurements \cite{PPGCMD2} are
based on the full SND data set which contains the largest statistics
of $\pi^0\pi^0\gamma$ events.
The  shape of the $\pi^0\pi^0$ mass
spectrum is described well by the sum of $\rho\pi$
and $f_0\gamma$ contributions. 
The contribution of $\sigma\gamma$
intermediate state is not required for description of the data
but is not completely excluded by our data.
It is shown that point-like transition model does not
fit the experimental  mass spectrum.
At the same time, models with intermediate kaon loops
reproduce the data. Predictions of the four-quark MIT bag model
of the $f_0$ meson are in a good agreement with
our results but we cannot exclude  other explanations.
Further theoretical studies and new data 
are needed to obtain consistent description of
the $f_0(980)$-meson quark structure and to detect
$\sigma\gamma$ contribution in the $\phi\to\pi^0\pi^0\gamma$ decay.

\section{Acknowledgments}

Authors are grateful to N.N.Achasov for 
useful discussions and valuable comments.

%\onecolumn
\newpage

\begin{table}
\caption{The matrix of detection efficiencies for
 three classes of events for the PHI96 experiment.  
}
\vspace {2pt}
\label{T96}
\begin{tabular}{lll}
\hline
  $j$  &  1  &  2   \\
\hline
 $m_{\pi\pi},~\mbox{MeV}$ & 300--700 & 700--1000  \\ 
\hline
$\epsilon^{1j}_{\pi\pi\gamma}$,~\% & 13.27 & 0.82   \\
$\epsilon^{2j}_{\pi\pi\gamma}$,~\% & 1.46 &  18.25   \\
$\epsilon^{3j}_{\pi\pi\gamma}$,~\% & 5.11 & 0.45   \\
\hline
\end{tabular}
\end{table}
\begin{table}
\caption{The matrix of detection efficiencies for
 three classes of events for the PHI98 experiment.  
}
\vspace {2pt}
\label{T98}
\begin{tabular}{lll}
\hline
  $j$  &  1  &  2   \\
\hline
 $m_{\pi\pi},~\mbox{MeV}$ & 300--700 & 700--1000  \\ 
\hline
$\epsilon^{1j}_{\pi\pi\gamma}$,~\% & 14.34 & 0.87   \\
$\epsilon^{2j}_{\pi\pi\gamma}$,~\% &  1.79 & 18.87   \\
$\epsilon^{3j}_{\pi\pi\gamma}$,~\% &  4.19 & 0.47   \\
\hline
\end{tabular}
\end{table}
\begin{table}
\caption{The results of the fit of the cross sections.}
\vspace {2pt}
\label{TF}
\begin{tabular}{lll}
\hline
 Experiment & PHI96 & PHI98 \\ 
\hline
$B(\phi\to\pi^0\pi^0\gamma)\cdot 10^4,\;m_{\pi\pi}<700\mbox{MeV}$ &
     $0.258\pm 0.111$ & $0.055\pm 0.080$ \\ 
$B(\phi\to\pi^0\pi^0\gamma)\cdot 10^4,\;m_{\pi\pi}>700\mbox{MeV}$ &
     $0.960\pm 0.102$ & $1.087\pm 0.086$ \\ 
$B(\phi\to\pi^0\pi^0\gamma)\cdot 10^4,\;m_{\pi\pi}>900\mbox{MeV}$ &
     $0.582\pm 0.085$ & $0.545\pm 0.068$ \\
\hline
\end{tabular}
\end{table}
\begin{table}
\caption{The  $m_{\pi\pi}$ mass bins, average
mass inside the bin $m_i$, the non zero elements of
the efficiency matrix $\epsilon_{i,j}$, and the normalized 
experimental invariant
mass spectrum for the $\phi\to\pi^o\pi^o\gamma$ decay
after background subtraction and acceptance corrections.
Only statistical error are shown. Systematic uncertainty
is 4.5~\%.}
\vspace {2pt}
\label{T0}
\begin{tabular}{lllllll}
\hline
 N & $m_{\pi\pi},\mbox{MeV}$ & $m_i,\mbox{MeV}$ & $\epsilon_{i-1,i}$ &
$\epsilon_{i,i}$ & $\epsilon_{i+1,i}$ & 
   $\frac{dB(\phi\to\pi^o\pi^o\gamma)}{dm_{\pi\pi}}\cdot10^8~(\mbox{MeV}^{-1})$\\
\hline
1 & 300-400 & 350 &   -   & 0.953 & 0.047 &  $3.36\pm4.47$ \\
2 & 400-500 & 450 & 0.037 & 0.920 & 0.043 &  $1.01\pm4.67$ \\
3 & 500-600 & 550 & 0.043 & 0.900 & 0.056 &  $1.94\pm3.11$ \\
4 & 600-700 & 650 & 0.042 & 0.718 & 0.240 &  $6.26\pm2.00$ \\
5 & 700-720 & 710 & 0.058 & 0.660 & 0.281 &  $9.46\pm6.46$ \\
6 & 720-740 & 730 & 0.227 & 0.536 & 0.237 & $13.99\pm6.14$ \\
7 & 740-760 & 750 & 0.221 & 0.544 & 0.235 & $13.53\pm6.07$ \\
8 & 760-780 & 770 & 0.240 & 0.537 & 0.223 & $18.23\pm6.26$ \\
9 & 780-800 & 790 & 0.242 & 0.546 & 0.212 & $16.93\pm6.08$ \\
10 & 800-820 & 810 & 0.221 & 0.558 & 0.220 & $24.04\pm7.36$ \\
11 & 820-840 & 830 & 0.233 & 0.559 & 0.208 & $28.07\pm7.65$ \\
12 & 840-860 & 850 & 0.223 & 0.572 & 0.205 & $26.59\pm7.59$ \\
13 & 860-880 & 870 & 0.215 & 0.564 & 0.221 & $34.15\pm8.28$ \\
14 & 880-900 & 890 & 0.202 & 0.613 & 0.185 & $44.37\pm9.37$ \\
15 & 900-920 & 910 & 0.192 & 0.630 & 0.179 & $43.15\pm8.95$ \\
16 & 920-940 & 930 & 0.197 & 0.651 & 0.153 & $61.66\pm10.38$ \\
17 & 940-960 & 950 & 0.171 & 0.698 & 0.131 & $65.45\pm10.71$ \\
18 & 960-980 & 969 & 0.178 & 0.728 & 0.095 & $70.14\pm12.61$ \\
19 & 980-1000& 987 & 0.214 & 0.786 &  -    & $30.52\pm10.08$ \\
\hline
\end{tabular}
\end{table}

\end{document}